\documentclass{jltp}

\usepackage{graphicx} 

\title{Bose-Condensed Gases in a 1D Optical Lattice at Finite Temperatures}
\author{E. Arahata and T. Nikuni}
\address{Department physics, Faculty of science, Tokyo University of Science, \\
1-3 Kagurazaka, Shinjuku-ku, Tokyo 162-8601, Japan}
\runninghead{E.Arahata and T.Nikuni}{Bose-Condensed Gases in a 1D Optical Lattice at Finite Temperatures}
\begin{document}
\maketitle
\begin{abstract}
We study equilibrium properties of Bose-Condensed gases in a one-dimensional (1D) optical lattice at finite temperatures. We assume that an additional harmonic confinement is highly anisotropic, in which the confinement in the radial directions is much tighter than in the axial direction. We derive a quasi-1D model of the Gross-Pitaeavskii equation and the Bogoliubov equations, and numerically solve these equations to obtain the condensate fraction as a function of the temperature. 

\end{abstract}

\section{Introduction}
Recently ultracold atoms confined in optical lattices have attracted attention both theoretically\cite{EPJD27}$^-$\cite{PRA 72} and experimentally.\cite{PRL86}$^-$\cite{Sc293} In particular, the dipole-mode frequency\cite{Sc293} and damping\cite{PRA66} were measured by the direct observation of an oscillating atomic current in a 1D optical lattice at finite temperatures. Several papers have discussed finite-temperature properties of ultracold atoms in a 1D optical lattice in addition to a highly anisotropic harmonic confinement, in which the radial confinement is much tighter than the axial confinement. Most theoretical studies have concentrated on the first Bloch band using the Bose-Hubbard model, and have ignored the effect of radial excitations.\cite{PRA 70}$^{,}$ \cite{PRA 73} This is only a good approximation when $k_{\rm{B}}T\ll E_{\rm{R}}$ and $k_{\rm{B}}T\ll \hbar\omega_\bot$, where $\it{E}_{\rm{R}}$ is the recoil energy that is roughly the highest energy in the first Bloch band, and $\hbar\omega_\bot$ is the first excitation energy in the radial direction. In order to obtain more quantitative results that is applicable to the experiments of Refs.~6-8, however, it is important to consider thermal excitations in the radial direction, since the experiments do not always satisfy $k_{\rm{B}}T\ll E_{\rm{R}}$ and $k_{\rm{B}}T\ll \hbar\omega_\bot$.    \par
 In this paper, we study equilibrium properties of Bose-condensed gases in a 1D optical lattice at finite temperatures using Hartree-Fock-Bogoliubov-Popov (HFB-popov) approximation with taking into account the effect of radial excitations. We will show that the radial excitations are important in the thermodynamic properties at finite temperatures.

\section{Quasi 1D modeling of a trapped Bose gas}
Our system is described by the following Hamiltonian :
\begin{equation}
\hat{H}=\int d\textbf{r}\left\{\hat{\psi}^\dagger (\textbf{r})\left[-\frac{\hbar^2}{2m}\nabla ^2+V_{\rm{ext}}(\textbf{r})\right]\hat{\psi }(\textbf{r})+\frac{g}{2}\hat{\psi}^\dagger(\textbf{r})\hat{\psi}^\dagger (\textbf{r})\hat{\psi }(\textbf{r})\hat{\psi }(\textbf{r})\right\},\label{H}
\end{equation}
where $g=\frac{4\pi\hbar^2a}{m}$ is the coupling constant determined by the $s$-wave scattering length $a$. The external potential $V_{\rm{ext}}$ is given by
$V_{\rm{ext}}(\textbf{r})=V_{\rm{trap}}(\textbf{r})+V_{\rm{op}}(z),
V_{\rm{trap}}(\textbf{r})=\frac{m}{2}\left[\omega _\bot ^2(x^2+y^2)+\omega_z^2z^2\right], 
V_{\rm{op}}(z)=sE_{\rm{R}}\cos^2(kz),
$
where $\omega _\bot , \omega_z$ are the frequencies of the harmonic trap, $k=\frac{2\pi}{\lambda}$ is fixed by the wavelength of the laser field creating the 1D lattice, $E_{\rm{R}}\equiv\frac{\hbar^2k^2}{2m}$ is the recoil energy, and $s$ represents the intensity of the laser beam. 
In order to take into account the quasi 1D situation $\omega _\bot \gg \omega _z $, we expand the field operator in terms of the radial wave function :
\begin{equation}
\hat{\psi }(\textbf{r})=\sum _{\alpha }\hat{\psi }_{\alpha}(z)\phi _\alpha (x,y)\label{expand},
\end{equation}
where $\phi_\alpha(x,y) $ is the eigenfunction of the radial part of the single-particle Hamiltonian,
\begin{eqnarray}
&&\left[-\frac{\hbar^2}{2m}\nabla _{\bot}^2+\frac{m}{2}\omega _\bot ^2(x^2+y^2)\right]\phi _\alpha (x,y)=\epsilon _\alpha\phi _\alpha (x,y)\label{xy1},
\end{eqnarray}
which satisfy the orthonormality condition
\begin{eqnarray}
 \int dxdy \ \phi ^\ast _\alpha (x,y)\phi _\beta  (x,y)=\delta _{\alpha \beta}\label{xy2}.
\end{eqnarray}
Here $\alpha=(n_x,n_y)$ is the index of the single-particle state with the eigenvalue $\epsilon_{(n_x,n_y)}=\hbar\omega_\bot(n_x+n_y+1)$.
Inserting (\ref{expand}) into (\ref{H}) and using (\ref{xy1}) and (\ref{xy2}), we obtain
\begin{eqnarray}
\hat{H}=\sum _\alpha \int dz \hat{\psi }^\dag_{\alpha}(z)\left[-\frac{\hbar^2}{2m}\frac{\partial ^2}{\partial z^2}+\frac{m}{2}\omega _z^2z^2+V_{\rm{op}}(z)+\epsilon _\alpha \right]\hat{\psi }_{\alpha}(z)\nonumber\\
+\sum _{\alpha  \alpha ^\prime \beta \beta^\prime}\frac{g_{\alpha  \alpha ^\prime \beta \beta^\prime}}{2}\int dz \hat{\psi }_{\alpha}^\dagger \hat{\psi }_{\beta}^\dagger \hat{\psi }_{\beta ^\prime}\hat{\psi }_{\alpha ^\prime},
\end{eqnarray}
where the renormalized coupling constant is defined by
\begin{equation}
g_{\alpha  \alpha ^\prime \beta \beta^\prime}\equiv g\int dxdy \phi _{\alpha}^\ast \phi _{\beta }^\ast \phi _{\beta ^\prime}\phi _{\alpha ^\prime}.
\end{equation}
\par Following the procedure described in Ref.~9
, we separate out the field operator $\hat{\psi}_\alpha$ into the condensate wavefunction and the noncondensed part;
$
\hat{\psi}_\alpha=\langle\hat{\psi}_\alpha\rangle+\tilde{\psi}_\alpha \equiv \Phi_\alpha+\tilde{\psi}_\alpha .
$
Within the HFB-Popov approximation, we obtain the generalized Gross-Pitaevskii (GP) equation,
\begin{eqnarray}
\mu \Phi _\alpha&=&\left[-\frac{\hbar^2}{2m}\frac{\partial ^2}{\partial z^2}+\frac{m}{2}\omega _z^2z^2+V_{\rm{op}}+\epsilon _\alpha \right]\Phi _\alpha\nonumber\\
&&+\sum _{\alpha ^\prime \beta \beta^\prime}g_{\alpha  \alpha ^\prime \beta \beta^\prime}\left(\Phi^\ast _{\beta}\Phi _{\beta^\prime}\Phi _{\alpha^\prime}+2\Phi _{\beta^\prime} \left\langle\tilde{\psi}_{\beta}^\dag \tilde{\psi}_{\alpha ^\prime}\right\rangle  \right).
\label{GPal}
\end{eqnarray}
The Popov approximation neglects the anomalous correlation  $\left\langle \tilde{\psi}\tilde{\psi}  \right\rangle$.\cite{PRB53} From the numerical solutions of the GP equation (\ref{GPal}) using the trap frequencies given later, we find that $|\Phi_\alpha|^2/|\Phi_0|^2 \ll 10^{-6}$ ($\alpha\neq 0$), and thus we will henceforth approximate $\Phi_\alpha\approx\Phi\delta_{\alpha,0}$, where we have denoted the lowest radial state as $\alpha=0=(0,0)$.\par
Taking the usual Bogoliubov transformations for the noncondensate,
\begin{eqnarray}
\tilde{\psi}_\alpha(z)=\sum_{j}\left(u_{j\alpha}\hat{\alpha}_j-v_{j\alpha}^*\hat{\alpha}_j^\dag  \right),
\end{eqnarray}
 we obtain the coupled Bogoliubov equations,
 \begin{eqnarray}
 L_\alpha u_{j \alpha}+\sum_{\alpha^\prime} \Biggl[ \left( 2g^\alpha_{\alpha^\prime} n_0 +g_{\alpha\alpha^\prime \beta \beta^\prime}\tilde{n}_{\beta\beta^\prime} \right) u_{j \alpha^\prime}-g^\alpha_{\alpha^\prime} n_0 v_{j \alpha^\prime} \Biggr]=E_j u_{j \alpha},\label{U}\\
  L_\alpha v_{j \alpha}+\sum_{\alpha^\prime} \Biggl[ \left( 2g^\alpha_{\alpha^\prime} n_0 +g_{\alpha\alpha^\prime \beta \beta^\prime}\tilde{n}_{\beta\beta^\prime}\right) v_{j \alpha^\prime}-g^\alpha_{\alpha^\prime} n_0 u_{j \alpha^\prime} \Biggr]=-E_j u_{j \alpha},\label{V}
 \end{eqnarray}
 where,
  \begin{eqnarray}
   L_\alpha&\equiv&-\frac{\hbar^2}{2m}\frac{\partial ^2}{\partial z^2}+\frac{m}{2}\omega _z^2z^2+V_{\rm{op}}(z)+\epsilon _\alpha-\mu ,\\
  n_0(z)&=&|\Phi(z)|^2,\\
   g^\alpha_{\alpha^\prime}&=&g_{\alpha\alpha^\prime00},\\
 \tilde{n}_{\beta\beta^\prime} &=&\left\langle\tilde{\psi}^\dag_\beta\tilde{\psi}_{\beta^\prime}  \right\rangle .
 \end{eqnarray} 
 Sums over the repeated indices $\beta, \beta^\prime$ are implied in Eqs.~(\ref{U}), and (\ref{V}).
 These equations define the quasi-particle excitation energies $E_j$ and the quasi-particle amplitudes $u_j$ and $v_j$.  
 Using the solutions of Eqs.~(\ref{U}) and (\ref{V}), one can obtain the noncondensate density from 
\begin{eqnarray}
\tilde{n}=\sum_{j \alpha\beta}\biggl[\left(u_{j\alpha}u_{j\beta}+v_{j\alpha}v_{j\beta}\right)N(E_j)+v_{j\alpha}v_{j\beta}\biggr],
\end{eqnarray} 
where $N(E_j)=1/\left[ \exp(\beta E_j )-1\right]$.
\section{Result}
Our calculation procedure is summarized as follows. Eq.~(\ref{GPal}) is first solved self-consistently for $\mu$ and $\Phi$ neglecting the interaction terms. Once $\Phi$ is known, $E_j$, $u_{j\alpha}$ and $u_{j\beta}$ are obtained from Eqs.~(\ref{U}) and (\ref{V}) with $\tilde{n}_{\beta\beta^\prime}$ set to zero. This is inserted into Eq.~(\ref{GPal}) and the process is repeated until convergence is reached. At each step, we define $\int dz \Phi^2= N-\tilde{N}$, where $\tilde{N}=\int dz \tilde{n}$ is the total number of noncondensate atoms.\par
Throughout this paper we use the following parameters:\cite{PRA66}   $m(^{87}$Rb)=1.44$\times 10^{-25}$ kg, $\omega_z/2 \pi= 9.0$ Hz, $\omega_{\bot}/2\pi= 92$ Hz, scattering length $a=5.82\times 10^{-9}$ m and the wavelength of the optical lattice $\lambda$=795 nm. We fix the total number of atoms as $N=4\times 10 ^5$.
\begin{figure}[htbp]
\centerline{\includegraphics[height=2.5in]{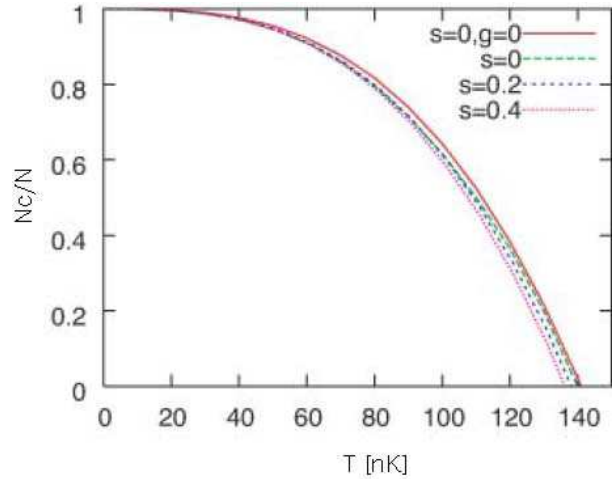}}
\caption {The condensate fraction $N_{\rm{c}}/$$N$ as a function of the temperature. $``s$=0, $g$=0" represents the ideal Bose gas result without a lattice potential given by $N_{\rm{c}}/N=1-(T/T^0_{\rm{c}})^3$, where $T^0_{\rm{c}}=$141 nK. 
$s$=0, 0.2, and 0.4 are calculated using the HFB-popov equation.}
 \label{fig:Tc}
\end{figure}

\begin{figure}[htbp]
\centerline{\includegraphics[height=2.5in]{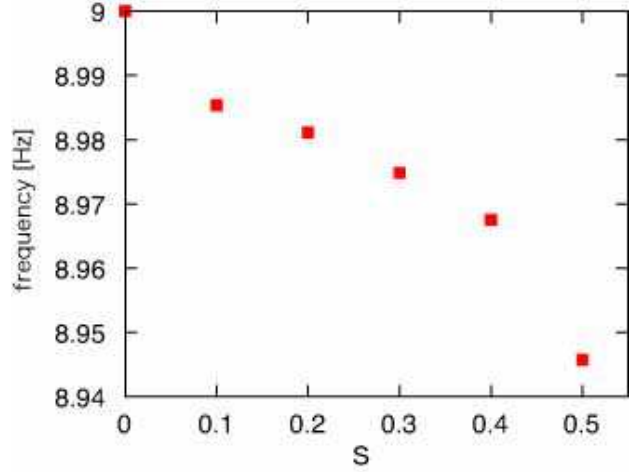}}
 \caption{The dipole mode frequency as a function of  the lattice depth $s$ with a fixed temperature $T=40$nK.}
  \label{fig:w}
\end{figure} 
In Fig.~\ref{fig:Tc}, we plot the condensate fraction $N_{\rm{c}}/$$N$ as a function of the temperature for various values of the lattice depth $s$.
It can be seen that each line falls to zero at approximately 140 nK, which is close to  the semiclassical prediction of the BEC transition temperature of an ideal Bose gas in a 3D harmonic trap $T_{\rm{c}}^0$$= 0.94\hbar (\omega_\bot^2\omega_z)^{1/3}N^{1/3}/k_{\rm{B}}$= 141 nK.    In contrast,  $T_{\rm{c}}$ of an ideal Bose gas in a 1D harmonic trap is $T_{\rm{c}}$$ \sim \hbar\omega_z\frac{N}{k_{\rm{B}}\ln N}$=13.4 $\mu$K. This shows that one must explicitly take into account the excitations in the radial direction in order to obtain the correct thermodynamic behavior at finite temperatures. \par
The highest $T_{\rm{c}}$ is obtained for an ideal gas without a lattice potential. The transition temperature deceases with increasing the lattice depth. This is due to a reduction of the excitation energy with increasing  the lattice depth. The energy shift can be seen from the dipole-mode frequency shift. In Fig.~\ref{fig:w}  we plot the dipole-mode frequency as a function of the lattice depth $s$ with a fixed temperature $T=40$nK. One can see that the frequency shift  from $s$=0 to 0.2 is bigger than the shift from $s$=0.2 to 0.4. This is consistent with the fact that $T_{\rm{c}}$ shift from $s$=0 to 0.2 is bigger than that from $s$=0.2 to 0.4.\par
\begin{figure}[htbp]
\centerline{\includegraphics[height=2.5in]{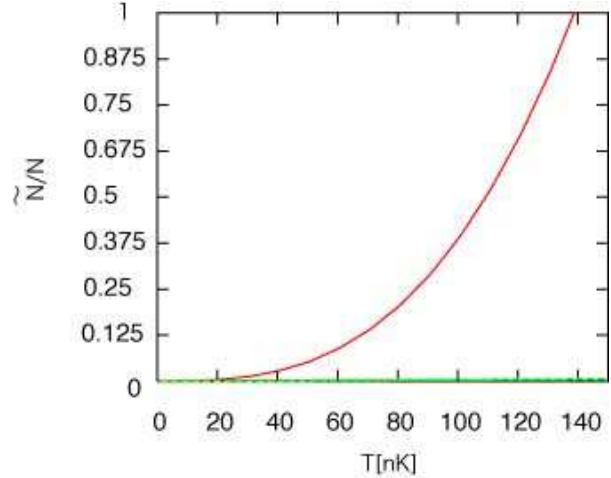}}
\caption{The temperature dependence of the noncondensate fraction for $s$=0.2. The solid line is the result including the radial excitations, while the dashed line is the result neglecting the radial excitations. }
  \label{fig:TildeN}
\end{figure}

Fig.~\ref{fig:TildeN} shows the effect of the radial excitations in the noncondensate fraction. The solid line shows the noncondensate fraction, while the dash line represent the result obtained by neglecting the radial excitations. one can clearly see that it is necessary to take into account the excitations in the radial direction at almost all temperatures. 
\section{Conclusion}
We have presented a detailed calculation of the condensate fraction in a 1D optical lattice at finite temperatures using the HFB-Popov approximation.
We explicitly included the radial excitations. Although the condensate is well described in terms of the lowest radial mode, it is necessary to take into account the thermal excitations in the radial direction at almost all temperatures. 
The shift of $T_{\rm{c}}$ with increasing the lattice depth is due to the shift of the low-energy excitations.    
In a future study, we will extend the present work to calculate damping rate of collective modes and study its dependence on the lattice depth $s$.  



\begin{thebibliography}{9}
\bibitem{EPJD27}\it{M. Kr\"{a}mer et al.,} \it{Eur. Phys. J. D} {\bf 27}, 247 (2003)
\bibitem{PRA 70}S. Tsuchiya and A. Griffin, \it{Phys. Rev. A } {\bf 70}, 023611(2004)
\bibitem{PRA 73}B. G. Wild, et al., \it{Phys. Rev. A} {\bf 73}, 023604(2006)
\bibitem{PRA 72} A. M. Rey, et al., \it{Phys. Rev. A} {\bf 72}, 033616(2005)     
\bibitem{PRL86}S. Burger et al., \it{Phys. Rev. Lett.} {\bf 86}, 4447(2001)
\bibitem{PRL94}C. D. Fertig et al., \it{Phys. Rev. Lett.} {\bf 94}, 120403(2005)
\bibitem{PRA66}F. Ferlaino et al., \it{Phys. Rev. A } {\bf 66}, 011604(2002)
\bibitem{Sc293}F. S. Cataliotti et al., \it{Science} {\bf  293}, 843(2001)
\bibitem{PRB53}A. Griffin,\it{ Phys. Rev. B} {\bf 53}, 9341(1995)

\end{thebibliography}
\end{document}